\documentclass[twocolumn,showpacs,preprintnumbers,amsmath,amssymb,aps]{revtex4}
\usepackage{graphicx}
\begin{document}

\title{Orbital Angular Momentum Exchange in an Optical Parametric Oscillator}
\author{M. Martinelli$^1$, J. A. O. Huguenin$^2$, P. Nussenzveig$^1$, A. Z. Khoury$^{2*}$}
\affiliation{1- Instituto de F\'\i sica - Universidade de S\~ao Paulo \\
PO Box 66318 CEP 05315-970  S\~ao Paulo-SP Brazil}
\affiliation{2- Instituto de F\'\i sica - Universidade Federal Fluminense \\
BR 24210-340 Niter\'{o}i-RJ Brazil}

\begin{abstract}
We present a study of orbital angular momentum transfer from pump 
to down-converted beams in a type-II Optical Parametric Oscillator. 
Cavity and anisotropy effects are investigated and demostrated to 
play a central role in the transverse mode dynamics. 
While the idler beam can oscillate in a Laguerre-Gauss mode, 
the crystal birefringence induces an astigmatic effect in the 
signal beam that prevents the resonance of such mode. 
\end{abstract}

\pacs{42.50.Dv; 42.65.Yj; 42.30.-d; 42.50.Lc}
\maketitle


\section{Introduction}

Early experiments have shown that circularly polarized light carries 
angular mometum \cite{A}. In a quantum description of light, this angular 
momentum is associated with the spin of the photon. More recently 
a significant attention has been given to the study of the orbital angular 
momentum of light, associated with phase singularities in the wavefront. 
In a paraxial description of wave propagation, it is found that 
Laguerre-Gaussian beams carry orbital angular momentum. Such beams 
can be experimentally produced either through astigmatic mode conversion 
with cylindrical lenses \cite{cillens,cillens2} or through holographic techniques 
\cite{holog1,holog2,moire}. While the holographic techniques are simpler, 
the astigmatic mode converters can operate with high power.

It has been demonstrated that parametric down conversion with free propagating 
beams conserves the Orbital Angular Momentum (OAM) of the pump beam. In the 
quantum optical domain this was observed through intensity correlation 
measurements, where twin-photon coincidences were obtained whenever the 
conservation condition was fulfilled \cite{amair}. In the classical counterpart, 
a stimulating beam was introduced and parametric amplification was shown to 
be conditioned to OAM conservation \cite{stimulated}.
Similar effects have already been observed in second harmonic generation 
\cite{shg1,shg2}.

So far, little attention has been given to OAM conservation in intracavity 
nonlinear coupling. Many studies have been done with transverse multimode OPOs, 
showing interesting possibilities in pattern formation and quantum images for 
cavities with degenerate transverse modes, like planar 
\cite{GattiLugiato,LugiatoGatti} and spherical cavities 
\cite{SphericalOPO,LugiatoMarzoli}. 
Moreover, experiments have shown pattern formation in confocal \cite{ConfocalSqz} 
and concentric \cite{Mathias} cavities, and oscillation in modes with higher order 
than the fundamental are common in many different experiments 
\cite{Suret,SaraDucci}. 

Apart from some theoretical studies on generation of phase singularities with 
nonlinear optical effects \cite{vortex1,vortex2}, only a few experimental results have 
been published in this subject, and to our knowledge, there is no result showing the 
necessary conditions for intracavity OAM transfer from the pump to the down-converted 
beams.
In the present work, we study the OAM transfer in a non-degenetrate, type-II 
Optical Parametric Oscillator (OPO). We show the conditions that must be 
satisfied for the OAM transfer, allowing one of the down-converted beams to 
oscillate with the same phase singularity of the pump beam. As we shall see, 
the astigmatism caused by the crystal birrefringence plays a central role in 
the selection of the beam oscillating in the Laguerre-Gauss mode.

\section{Experimental Setup}

\begin{figure}[ht]
\includegraphics[clip=,width=8cm]{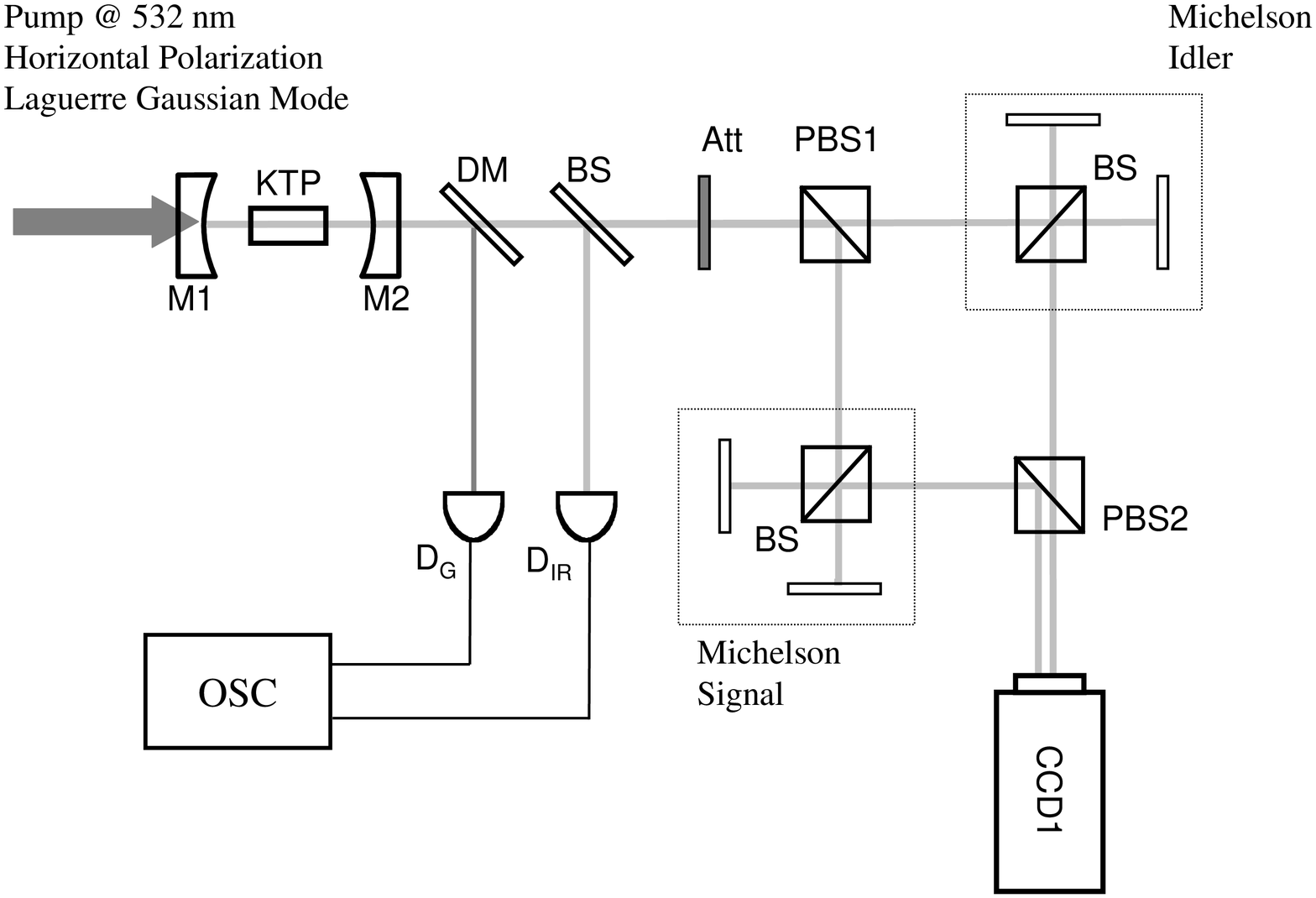}
\caption{\label{setup}Setup for the study of the phase singularities in the output of the 
type-II triply resonant OPO. The mode converter introduced in the pump beam is not shown}
\end{figure}

The experimental setup is shown in Fig.\ref{setup}. The OPO is made by two 
spherical mirrors M1 and M2, with equal curvatures $R_m=13 mm$. Inside the cavity, 
we have a KTP crystal (by Cristal Laser) 10 mm long, cut for noncritical phase 
matching in 532 nm to 1064 nm down conversion at room temperature. In this case, 
the crystallographic axes $(x,y,z)$ of the cristal are oriented as follows. 
The $z$ axis of the crystal is vertically oriented while the propagation direction 
lies on the horizontal plane ($xy$). The $x$ axis forms an angle $\phi=23.5 ^\circ$ 
with respect to the propagation direction. 

The mirrors have high reflectance for the infrared (R = 99.8 \% @ 1064 nm), and a small 
transmition at the pump wavelength (R = 92 \% @ 532 nm). Crystal losses in the infrared 
comes mainly from surface reflection, reduced by anti-reflective coating (R = 0.1 \%), 
since crystal absorption at this wavelength  is small (0.05 \%). For the pump, we have 
reflection losses (R = 0.5 \%) and crystal absorption, increased by gray-tracking effects 
\cite{graytracking}.
The cavity length is controlled by a piezo actuator on the mirror, and the cavity is kept 
nearly confocal, in order to help the alignment and reduce the consequences of the walk-off.

\begin{figure}[ht]
\vspace*{1.5cm}
\includegraphics[clip=,width=8cm]{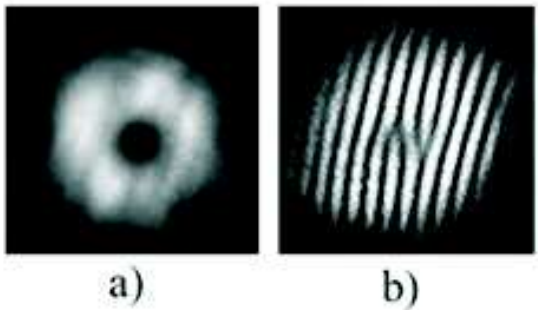}
\caption{\label{pump}a) Transverse profile of the pump; b) Interference pattern showing the 
topological defects characteristic of phase singularities}
\end{figure}

The OPO is pumped by the second harmonic of a Nd:YAG laser (Lightwave 142). This laser 
generates a TEM$_{00}$ gaussian beam, that is converted to a nearly Hermite-Gauss 
TEM$_{01}$ beam \cite{Petrov}. With a telescope formed by two cylindrical lenses, we implemented 
a mode converter that produces a Laguerre-Gauss beam \cite{cillens}, with a good cylindrical 
simetry for the intensity and a phase singularity in the center. This phase singularity was 
evidenced by the self interference pattern obtained in a Michelson interferometer. 
In Fig.\ref{pump} we show the transverse profile and interference pattern of the beam used 
to pump the OPO. The resulting pump power is 60 mW. The beam is horizontally polarized, and 
mode matched to the cavity with the help of coated lenses.
Although the mirrors were high reflecting at 1064 nm, the output power coming out from the 
cavity through M2 can be detected by a PIN photodiode. The green light coming from the cavity 
is filtered by a dichroic mirror (DM), and detected by an amplified Si photodetector (D$_G$). 
The infrared light is detected by a PIN InGaAs photodiode D$_{IR}$ (ETX-300, from Epitaxx), 
that samples part of the output beam that is reflected by a beam splitter (BS). 

In the output, signal and idler beams are separated by a polarizing cube (PBS). Adopting the usual 
convention in type-II OPOs, the idler beam polarization is horizontal, and the signal beam has 
a vertical polarization, alligned to the crystal $z$ axis. Each down converted beam is sent into 
a Michelson interferometer made by a nonpolarizing 50/50 beam splitter (BS) and two flat mirrors, 
in order to produce interference fringes that can reveal the existence of a phase singularity. 
The two outputs are recombined in another polarizign cube (PBS) and sent onto a CCD camera, that 
is used to register either the interference pattern or the intensity profile of the beam.

\begin{figure}[ht]
\includegraphics[clip=,width=8cm]{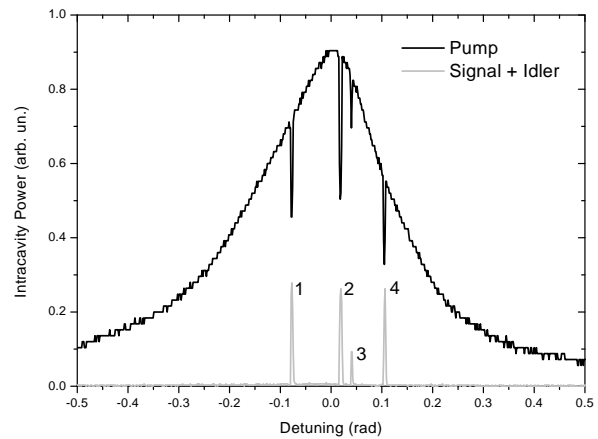}
\caption{\label{peaks}Resonance peak for the pump beam, showing the oscillation of the OPO. 
The infrared peaks are labeled in order to identify the different images shown in 
Fig.\ref{image}}
\end{figure}

\begin{figure}[ht]
\includegraphics[clip=,width=8cm]{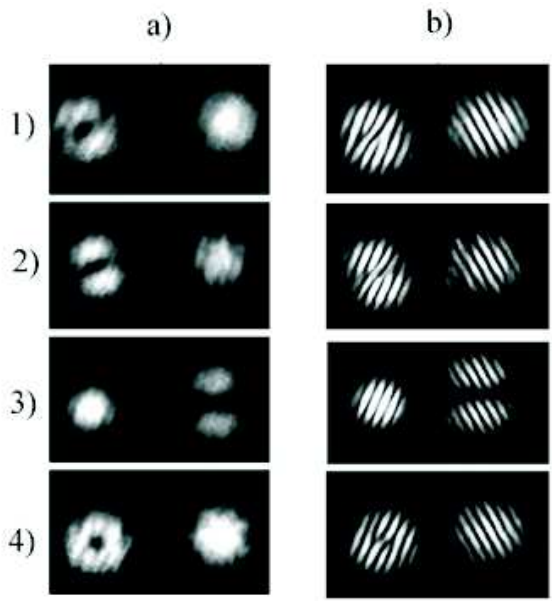}
\caption{\label{image}a) Intensity patterns for signal (right) and idler (left) beams 
labeled in correspondence to the infrared peaks shown in Fig.\ref{peaks}. 
b) Self interference patterns showing the presence or not of phase singularities.}
\end{figure}

The output power of the pump and infrared beams is measured as the cavity length is scanned. 
The corresponding resonance peaks are shown in Fig.\ref{peaks}. A wide peak is obtained for 
the pump, over which narrow dips appear due to the pump depletion in different oscillation 
regimes. The resonance peaks for the infrared are also shown in Fig.\ref{peaks}. They coincide 
with the depletion dips in the pump resonance. 
Expanding the curve, we can observe that the depletion dips have a 
parabolic shape, in good agreement with the depletion expected for a triply resonant OPO 
\cite{Debuisschert}. From the finesse of the resonance peak for the pump,  we measure 
29 \% of internal losses in the cavity. For signal and idler modes, the fitting of the 
parabolic depletion gives a value of 1 \% for the infrared losses. The threshold power 
for parametric oscillation is around 20 mW.
The OPO could be held in the oscillating peaks, and it would rest with a continuous output 
for as long as 10 minutes. In this situation, we registered the output image of signal 
and idler beams, as well as their self interference patterns. These images are shown in 
Fig.\ref{image}. They are lebeled in correspondence to the oscillation peaks shown in 
Fig.\ref{peaks}.
In images 1 and 4, the output intensity in the idler is the one of a Laguerre-Gauss beam. 
The corresponding interference patterns show the topological defects in the center of the 
Laguerre-Gauss beam characteristic of phase singularities. In this situation, the idler 
beam carries the orbital angular momentum of the down converted pump photons. In image 2, 
the shape of the idler beam is intermediary between a first order Laguerre-Gauss 
and a diagonal first order Hermite-Gauss modes. A vortex can still be observed through the 
interference fringes. In both cases, the signal beam remains in the fundamental gaussian mode.
Following the Poincar\'e-sphere representation proposed in Ref.\cite{poincare}, we can look at 
the idler mode shown in image 2 as an orbital equivalent of an elliptical polarization. 

An interesting effect appears in image 3. In this situation, the signal beam oscillates in 
the transverse mode with higher order, but with no angular momentum. The output is a pure 
Hermite-Gauss TEM$_{01}$ mode, vertically oriented, while the idler remains in the fundamental 
gaussian mode. Therefore, the orbital angular momentum is not conserved in the parametric down 
conversion process, and the crystal is expected to suffer a twisting torque. 
This effect is analogous to the mechanical torque applied to a quater waveplate used for 
light polarization conversion \cite{A}, or to a pair of cylindrical lenses used for 
transverse mode conversion \cite{cillens}.

The reason for this assimetry in the OAM conservation of the pump can be explained when 
the propagation of paraxial beams in anisotropic media is investigated and the 
effect of cavity mode selection is considered.

\section{Paraxial waves in a birrefringent crystal}

In a type-II down conversion, we use the crystal birrefringence to achieve the 
desired phase matching condition \cite{Dmitriev,Yariv}. It was shown by many 
authors that the paraxial equation, from which we can derive the propagation modes 
of a beam in free space or isotropic medium, will change when we work with an 
anisotropic medium \cite{Mejias,Ciattoni}.
Here we will extend the study of Fleck and Feit \cite{FleckFeit} of paraxial propagation 
in uniaxial crystals to the biaxial case, adapting their description to the case of a 
crystal inside a cavity. Our aim is to reduce the wave equations to the paraxial 
wave equations that define the Hermite-Gauss modes coupled to the resonances of a 
linear cavity. 

Let us define the crystalographic axis as $(x,y,z)$.
The KTP crystal used in our experiment is a quasi-uniaxial one with 
$n_x\sim n_y\neq n_z$, where $n_{x(y,z)}$ is the refraction index for $x(y,z)$ polarized 
light.  
Since the displacement vector $\mathbf{D}$ satisfies 
$\nabla\cdot\mathbf{D}=0$, we can write the wave equation for the electric field 
$\mathbf{E}$, derived from Maxwell's equations, as
\begin {equation}
\nabla^2\mathbf{E}-
\nabla\,\left(\nabla\cdot\mathbf{E}-\frac{\nabla\cdot\mathbf{D}}{\alpha}\right)
+k_0^2\,\mbox{\boldmath $\varepsilon$}\cdot\mathbf{E}=0,
\label{eqonda}
\end{equation}
where $k_0=\omega/c$ is the wave number in vacuum corresponding to frequency $\omega$, 
and $\alpha$ is a constant to be conveniently chosen. This constant will significantly 
simplify the paraxial propagation analysis in the birrefringent medium. 
The constitutive relation $\mathbf{D}=\mbox{\boldmath $\varepsilon$}\cdot\mathbf{E}$ depends 
on the dieletric tensor $\mbox{\boldmath $\varepsilon$}$, that is diagonal when we use 
the crystalographic coordinates
\begin{equation}
\mbox{\boldmath $\varepsilon$} = \left[\begin{array}{ccc}
n_{x}^2 & 0 & 0 \\
0 & n_{y}^2 & 0 \\
0 & 0 & n_{z}^2
\end{array}\right].
\end{equation}

\begin{figure}[ht]
\includegraphics[clip=,width=8cm]{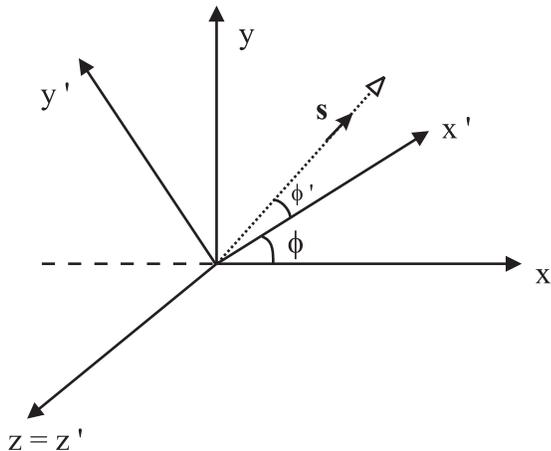}
\caption{\label{coords}Coordinate system used to describe the paraxial propagation 
through the anisotropic crystal. The walk-off angle $\phi^{\prime}$ is indicated as 
the angle between the Poynting vector ${\mathbf S}$ and the propagation axis 
$x^{\prime}\,$.}
\end{figure}

The wave equations for the electric field components can be derived from Eq.(\ref{eqonda})  
by using the constitutive relation and choosing $\alpha=n_y^2$ to obtain 
\begin{eqnarray}
& &\frac{n_x^2}{n_y^2}\,\partial_x^2 E_x\,+\,\partial_y^2 E_x\,+\,\partial_z^2 E_x 
-\left(1-\frac{n_z^2}{n_y^2}\right)\,\partial_x\partial_z E_z 
\nonumber\\
&+&k_0^2\,n_x^2\,E_x=0\;,
\nonumber\\
& &\partial_x^2 E_y\,+\,\partial_y^2 E_y\,+\,\partial_z^2 E_y 
-\left(1-\frac{n_x^2}{n_y^2}\right)\,\partial_y\partial_x E_x
\nonumber\\ 
&-&\left(1-\frac{n_z^2}{n_y^2}\right)\,\partial_y\partial_z E_z + k_0^2\,n_y^2\,E_y=0\;,
\label{eqonda2}\\
& &\partial_x^2 E_z\,+\,\partial_y^2 E_z\,+\,\frac{n_z^2}{n_y^2}\,\partial_z^2 E_z 
-\left(1-\frac{n_x^2}{n_y^2}\right)\,\partial_z\partial_x E_x 
\nonumber\\
&+&k_0^2\,n_z^2\,E_z=0\;.
\nonumber
\end{eqnarray}
Notice that for an uniaxial crystal ($n_x=n_y$) we recover the equations obtained 
in ref.\cite{FleckFeit}. Let us now consider propagation along an axis $x^{\prime}$ 
lying on the $xy$ plane with an angle $\phi$ with respect to the crystalographic 
axis $x$, as shown in Fig.\ref{coords}. 
This definition of $\phi$ has the advantage to match the angle and axis 
definitions usually given by crystal manufacturers. For our KTP crystal, cut for 
type II phase matching of $532$ and $1064$ nm, we have $\phi=23.5^{\circ}$. 
A rotated reference frame $(x^{\prime},y^{\prime},z)$ can be used to describe the 
propagation inside the crystal. The coordinate transformation between the two 
frames is
\begin{eqnarray}
x^{\prime}&=&x\,\cos\phi + y\,\sin\phi\;,
\nonumber\\
y^{\prime}&=&-x\,\sin\phi + y\,\cos\phi\;,
\\
z^{\prime}&=&z\;.
\nonumber
\label{transf}
\end{eqnarray}

\subsection{Plane wave analysis}

Two orthogonally polarized plane wave solutions propagating along $x^{\prime}$ 
can be found for Eqs.(\ref{eqonda2}). 
One with $E_x=E_y=0$ and $E_z\neq 0$ ($z$ polarized) and another  
polarized in the $xy$ plane ($E_z=0$). 
For the $z$ polarized solution, only the last of Eqs.(\ref{eqonda2}) remain and its 
solution is 
\begin{equation}
E_z=E_{0z}\,{\rm e}^{i\,n_z k_0\,x^{\prime}}\;.
\label{pwz}
\end{equation}
The plane wave solution polarized on the $xy$ plane can be found by making $E_z=0$ 
in the first two of Eqs.(\ref{eqonda2}). 
This solution is of the kind 
\begin{equation}
{\mathbf E}={\mathbf E_0}\,\,{\rm e}^{i\,(k_x\,x+k_y\,y)}\,, 
\label{pwxy}
\end{equation}
where ${\mathbf E_0}=E_{0x}\,\hat{x} + E_{0y}\,\hat{y}$. From substitution of 
Eq.(\ref{pwxy}) in the first one of Eqs.(\ref{eqonda2}), we find 
\begin{equation}
\frac{k_x^2}{n_y^2}+\frac{k_y^2}{n_x^2}=k_0^2\,
\label{elipsek}
\end{equation}
which is the projection of the well known index ellipsoid on the $xy$ plane. 
By making $k_x=n\,k_0\,\cos\phi$ and  $k_y=n\,k_0\,\sin\phi$, we get
\begin{equation}
\frac{\cos^2\phi}{n_y^2}+\frac{\sin^2\phi}{n_x^2}=\frac{1}{n^2}\,
\label{elipsen}
\end{equation}
where $n$ is the index of refraction for propagation along $x^{\prime}$. 
On the other hand, if we substitute Eq.(\ref{pwxy}) in the constitutive relation 
$\mathbf{D}=\mbox{\boldmath $\varepsilon$}\cdot\mathbf{E}$ and in 
$\nabla\cdot{\mathbf D}=0$ we find that 
\begin{equation}
n_x^2\,k_x\,E_x + n_y^2\,k_y\,E_y = 0\;.
\end{equation}
Since $n_x\neq n_y$, this means that ${\mathbf E}$ and ${\mathbf k}$ are 
not orthogonal. Therefore, the Poynting vector ${\mathbf S}$, that is 
orthogonal to ${\mathbf E}$, is not parallel to ${\mathbf k}\,$. 
Let us call $\phi^{\prime}$ the angle between 
${\mathbf S}$ and ${\mathbf k}\,$.
A straightforward geometric analysis allows one to obtain a simple 
relation between $\phi$ and $\phi^{\prime}$: 
\begin{equation}
\tan\phi^{\prime}=\frac{\sin\phi\,\cos\phi\,(n_y^2 - n_x^2)}
{n_x^2\,\cos^2\phi + n_y^2\,\sin^2\phi}\;.
\label{angwalkoff}
\end{equation} 
This angle is represented in Fig.\ref{coords}. It 
is related to the well known walk-off effect, which appears as a 
consequence of the crystal anisotropy. However, as we shall see shortly, 
the $z$ polarized field will also have an anisotropic effect  
when the propagation of a transversely finite beam is considered. 
This effect appears as an astigmatic deformation of the beam during the 
propagation along the crystal.

\subsection{Paraxial propagation}

On ther other hand, to obtain a direct solution of eq.(\ref{eqonda}) for a 
paraxial beam propagation is not so straightforward and some carefull 
aproximations have to be made to uncouple the diferential equations for each 
polarization. 
For the $z$ component, the wave equation has the form
\begin{eqnarray}
& &\left(\partial^2_x + \partial^2_y + \frac{n_{z}^2}{\alpha}\partial^2_z + 
k_0^2\,n_{z}\right)E_{z} 
- \left(\frac{\alpha-n_{x}^2}{\alpha}\right)\partial_x\partial_z E_{x}
\nonumber\\ 
&-&\left(\frac{\alpha-n_{y}^2}{\alpha}\right)\partial_y\partial_z E_{y}=0\;.
\label{Ez}
\end{eqnarray}
To reduce this equation to the paraxial wave equation for $z$ polarization, we can 
begin by eliminating the terms with cross derivatives. One way to do this is to 
approximate the biaxial crystal by an uniaxial one for the $z$ polarization. 
This is valid since $|n_{x}-n_{y}|\ll|n_{z}-n_{y}|$.
If we chose $\alpha=n^2$, we have 
$\left|\frac{\alpha-n_{i}^2}{\alpha}\right|\cong 10^{-2}$ for $i=\{x,y\}$, 
giving a very small contribution. In the limit $n_{x}=n_{y}=n$ these terms 
will vanish, and we have the uniaxial crystal studied in ref.\cite{FleckFeit}.

A paraxial solution $E_{z}=u_{z}(x^{\prime},y^{\prime},z){\rm e}^{-i n_z k_0 x^{\prime}}$ 
of Eq.(\ref{Ez}) can be obtained if we adopt the rotated reference frame. 
The resulting equation is close to the paraxial wave equation, except for the assimetry 
in the coefficients of the transverse second-order derivatives:
\begin{equation}
\partial^2_{y^{\prime}} u_{z} 
+ \frac{n_{z}^2}{n^2}\,\,\partial^2_{z} u_{z} =
2\,i\,n_{z} k_0\,\partial_{x^{\prime}} u_{z} \;.
\label{paraxuz}
\end{equation}

The assimetry between the transverse coordinates $y^{\prime}$ and $z$ appears as a 
rescaling of the $z$ coordinate. This means that the optical beam follows an astigmatic 
propagation inside the crystal with diferent diffraction scales for each transverse 
coordinate. Let us separate the dependence of $u_{z}$ on $y^{\prime}$ and $z$ making 
$u_{z}(x^{\prime},y^{\prime},z)=U_{z}(x^{\prime},y^{\prime})\,V_{z}(x^{\prime},z)$, in order 
to obtain two paraxial wave equations for the beam diffraction in each transverse direction: 
\begin{eqnarray}
\partial^2_{y^{\prime}} U_{z}&=&2in_z k_0\,\partial_{x^\prime} U_{z}\;,
\nonumber\\
\frac{n_z^2}{n^2}\partial^2_z V_{z}&=&2in_z k_0\,\partial_{x{^\prime}} V_{z}\;.
\label{Zuzuy}
\end{eqnarray}
When calculating the propagation of the beam through an OPO cavity, this diffraction assimetry 
can be seen as a different effective length of the crystal for each transverse dependence of the 
mode function. For a crystal with length $\ell$, the effective length for the $U_{z}$ propagation 
will be $\ell/n_{z}$, as usual in the treatment of beam propagation through an uniform crystal
~\cite{Schwob,Yariv}. For $V_{z}$, the effective length it will be $\ell n_{z}/n^2$, resulting 
in an assimetry in the effective cavity length for each transverse evolution. The calculation of 
the cavity geometry, and the resulting beam parameters expressed by the Rayleigh length $x_R$, 
will therefore differ for the two transverse coordinates. 

Let us now turn to the paraxial solution for the field polarized on the $xy$ plane. 
Since a plane wave solution with $E_z=0$ can be found, it is natural to concieve a 
paraxial solution for which $E_z$ is negligeable. Therefore, if we choose 
$\alpha=n_{x}^2$ in Eq.(\ref{eqonda}) and use the rotated coordinates, we obtain the 
following propagation equation for $E_y$
\begin{eqnarray}
&&\left[\cos^2\phi \, \partial^2_{x^{\prime}} + \sin^2\phi \, \partial^2_{y^{\prime}}-
\sin2\phi \, \partial_{x^{\prime}}\partial_{y^{\prime}}\right. 
\nonumber\\
&&+ 
\frac{n_{y}^2}{n_x^2}\left(\sin^2\phi \, \partial^2_{x^{\prime}} +
\cos^2\phi \, \partial^2_{y^{\prime}}+ \sin2\phi \, \partial_{x^{\prime}}
\partial_{y^{\prime}}\right)
\nonumber\\ 
&&+\left. \partial^2_{z} + k_0^2\,n_{y}\right]E_{y} = 0
\label{rfEy}
\end{eqnarray}

We now try a paraxial solution of the form 
$E_{y}=u_{y}(x^{\prime},y^{\prime},z){\rm e}^{-i n k_0 x^{\prime}}$ 
in Eq.(\ref{rfEy}), using Eq.(\ref{elipsen}) and making the paraxial 
approximation to obtain:
\begin{eqnarray}
&&2ik_{0}n_{y}\left( \cos^2\phi + \frac{n_{y}^2}{n_{x}^2}\sin^2\phi \right)^{1/2}
\left( \partial_{x^{\prime}}u_{y} + \tan \phi^{\prime}\partial_{y^{\prime}}u_{y} \right)
= 
\nonumber\\
&&\left( \sin^2\phi + \frac{n_{y}^2}{n_{x}^2}\cos^2\phi \right)\partial^2_{y^{\prime}}u_{y}
+ \partial^2_{z}u_{y}
\label{paraxuy1}
\end{eqnarray}
where $\phi^{\prime}$ is the walk-off angle given by Eq.(\ref{angwalkoff}). 
In order to obtain a paraxial wave equation, a second coordinate transformation 
$y^{\prime\prime}= y^{\prime} - \tan\phi^{\prime}x^{\prime}$ is necessary. 
This transformation corresponds to a transverse offset of the $xy$ polarized 
beam. Using Eq.(\ref{elipsen}) and defining 
$\xi^2=\sin^2\phi + (n_{y}^2/n_{x}^2)\cos^2\phi\,$, 
we can rewrite Eq.(\ref{paraxuy1}) as
\begin{equation}
2ik_{0}\frac{n_{y}^2}{n}\,\partial_{x^{\prime}}u_{y} =
\xi^2\,\partial^2_{y^{\prime\prime}}u_{y}
+ \partial^2_{z}u_{y}\;,
\label{paraxuy2}
\end{equation}
that is, the usual paraxial equation with wave vector $k_0\,(n_y^2/n)$ and a rescaled 
transverse coordinate $y^{\prime\prime}/\xi\,$. 
However, since $n_x\sim n_y$, this transverse rescalling is 
much smaller than the one present in the $z$ polarized field. Therefore, while the 
$z$ polarization has a significant astigmatism but no walk-off, the $xy$ polarization 
presents walk-off and a small astigmatism. From now on we shall designate the 
$xy$ polarized field as the {\it extraordinary} wave and the $z$ polarized field as 
the {\it ordinary} wave.

As we made for the $z$ component, we can try a factorized solution of the paraxial
wave equation (\ref{paraxuy2}) of the form 
$u_{y}= U_{y}(x^{\prime},y^{\prime\prime})\,V_{y}(x^{\prime},z)\,$, so that 
\begin{eqnarray}
\xi^2\,\partial^2_{y^{\prime\prime}} U_{y}&=&
2ik_0\,(n_y^2/n)
\partial_{x^{\prime}} U_{y}\;,
\nonumber\\
\partial^2_{z} V_{y}&=&
2ik_0\,(n_{y}^2/n)\,\partial_{x^{\prime}} V_{y}\;.
\label{Yuzuy}
\end{eqnarray}
Thus, the paraxial propagation inside the crystal is well described by 
Eqs.(\ref{Zuzuy}) and (\ref{Yuzuy}) for the ordinary and extraordinary waves respectively.
A paraxial equation for the $x$ component of the extraordinary wave can be obtained 
on the same lines leading to Eq.(\ref{Yuzuy}).

\section{Astigmatic cavity}

All scalling parameters appearing in Eqs.(\ref{Zuzuy}) and (\ref{Yuzuy}) can be 
absorbed by a suitable definition of an effective wave number $k_{eff}$ for each 
transverse direction and for each polarization. This brings Eqs.(\ref{Zuzuy}) 
and (\ref{Yuzuy}) to the general form
\begin{equation}
\partial^2_{y} U(x,y)=2ik_{eff}\,\partial_{x} U(x,y)\; .
\label{paraxeff}
\end{equation}
The normalized solution of Eq.(\ref{paraxeff}) is \cite{siegman}
\begin{eqnarray}
&&U(x,y)=
\left(\frac{k_{eff}\,x_R}{\pi2^{2n}n!^2\left(x^2+x_R^2\right)}\right)^{1/4}
H_n\left(y\sqrt{\frac{k_{eff}\,x_R}{x^2+x_R^2}}\right)\times
\nonumber\\
&&\exp\left[-i\frac{k_{eff}\,y^2}{2(x+ix_R)}-i\left(n+\frac{1}{2}\right)
\arctan\left(\frac{x}{x_R}\right)\right]
\label{beameq}
\end{eqnarray}
where $x_R$ is the Rayleigh length, and $H_n(x)$ is the Hermite polinomial 
of order $n\geq0$. 
The term $\arctan{x/x_R}$ is the well known Gouy phase shift. 
This term avoids multiple resonances of high order Hermite-Gaussian (HG) modes 
in a high finesse cavity for the signal and idler modes of the OPO.
The beam propagation is characterized by the beam waist $w_0=\sqrt{2\,x_R/k_{eff}}$ 
and the wavefront curvature 
$R(x)=x\,\left(1+x_R^2/x^2\right)\;$.
The change in the effective wave number is equivalent (in terms of beam diffraction) 
to the propagation in a shorter length of free space. Since the effective wave number 
depends both on polarization and transverse direction, we can consider a different 
propagation length in each case. 

Let us now consider the refractive index of the KTP crystal at 1064 nm ($n_{x}=1.7404, 
n_{y}=1.7479, n_{z}=1.8296$) and 532 nm ($n_{x}=1.7797, n_{y}=1.7897, n_{z}=1.8877$), 
according to the manufacturer, Cristal Laser S.A. We have, for the extraordinary wave, 
a refractive index $n(1064 nm)=1.7467$ and $n(532 nm)=1.7881$.
From the distance $L_0=17.4 mm$ between the mirrors in our near-confocal cavity, 
and the crystal length $\ell=10.0 mm$, we can calculate the effective length $L$ of 
the cavity for each transverse mode, and for each polarization, in the infrared case. 
Using the relation
\begin{equation}
L=L_0\,-\,\ell\,\frac{k_{eff}-k_0}{k_{eff}}
\end{equation}
we obtain, from the values of $k_{eff}$ given by Eqs.\ref{Zuzuy},\ref{Yuzuy}
\begin{eqnarray}
L^{o}_{y^{\prime}}&=& 12.87 mm,\qquad
L^{o}_z= 13.40 mm,
\nonumber\\
L^{e}_{y^{\prime\prime}}&=& 13.17 mm,\qquad
L^{e}_z= 13.12 mm\; ,
\end{eqnarray}
where the superscript $o$($e$) refer to the ordinary (extraordinary) wave. 
The effect of the walk off for the extraordinary wave has been taken into account, 
but the correction was $\sim 10^{-4}$, and could be neglected. The values of the Rayleigh 
length inside the cavity ($x_R^2=L^2(2R_m-L)/4$) for each transverse direction of the beam, 
and for each polarization, differ by less than 1\%, and cannot be noticed in the 
free-propagating beam
\begin{eqnarray}
x_{Ry^\prime}^o&=&6.500 mm,\qquad
x_{Rz}^o= 6.497 mm,
\nonumber\\
x_{Ry^{\prime\prime}}^e&=&6.499 mm,\qquad
x_{Rz}^e= 6.500 mm \;.
\end{eqnarray}
On the other hand, the total Gouy phase shift accumulated in a round trip inside the cavity 
$\Phi=4\arctan(\sqrt{L}/\sqrt{2R_m-L})$ will be
\begin{eqnarray}
\Phi_{y^\prime}^o&=&3.122 rad,\qquad
\Phi_{z}^o= 3.204 rad,
\nonumber\\
\Phi_{y^{\prime\prime}}^e&=&3.167 rad,\qquad
\Phi_{z}^e= 3.161 rad.
\end{eqnarray}

The phase added in a round trip depends on the order of the Hermite-Gauss $TEM_{mn}$ mode 
resonating inside the cavity. The total Gouy phase for this mode is 
\begin{equation}
\Phi=(m+1/2)\Phi_z+(n+1/2)\Phi_y.
\label{phase}
\end{equation}
From the calculated values of the Gouy phase shift, we see that there will be a small phase 
diference between the TEM$_{01}$ and the TEM$_{10}$ modes. This difference will result in a 
splitting of the resonance position. At 1064 nm, this separation is of $82 mrad$ for 
the ordinary wave, and $6 mrad$ for the extraordinary one.

\begin{figure}[ht]
\includegraphics[clip=,width=8cm]{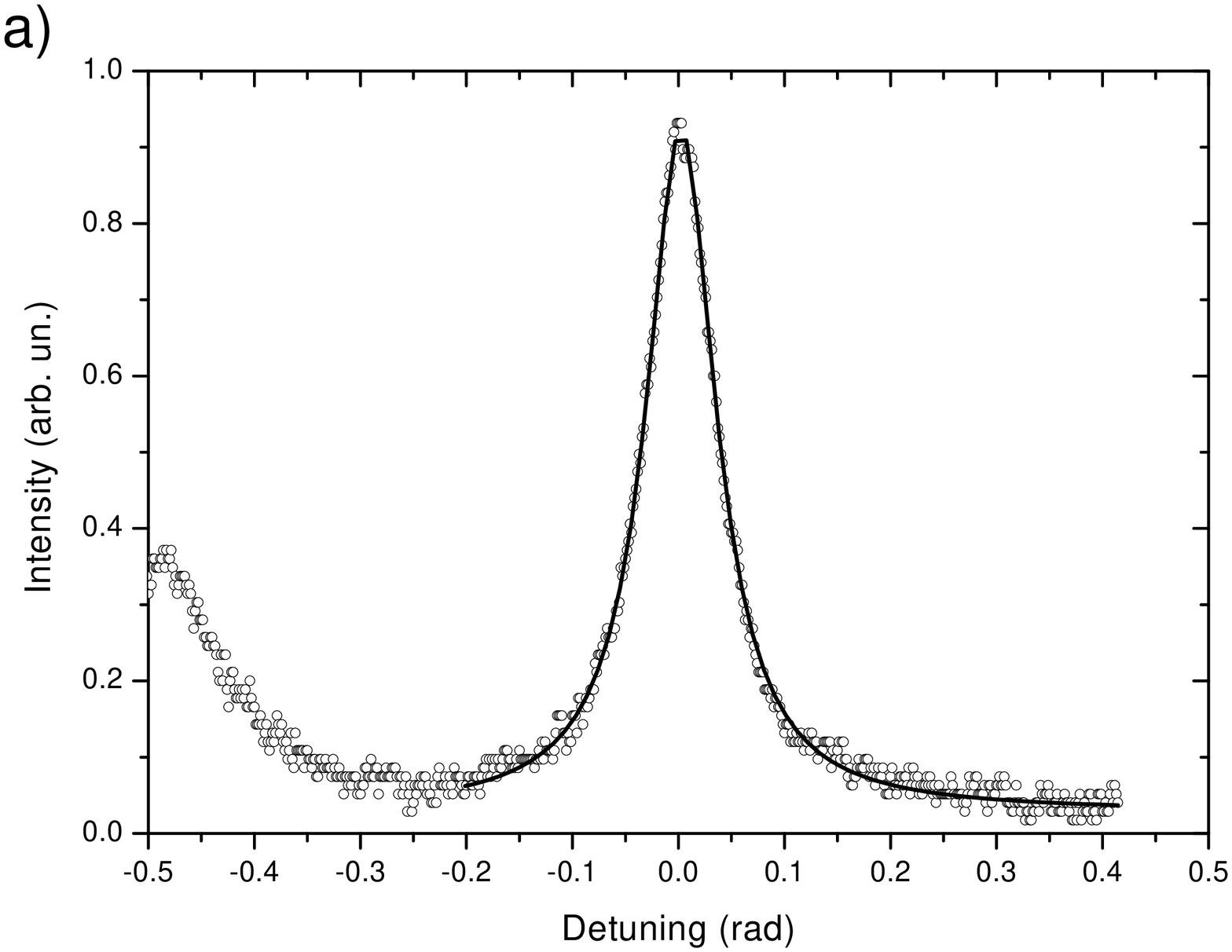}
\includegraphics[clip=,width=8cm]{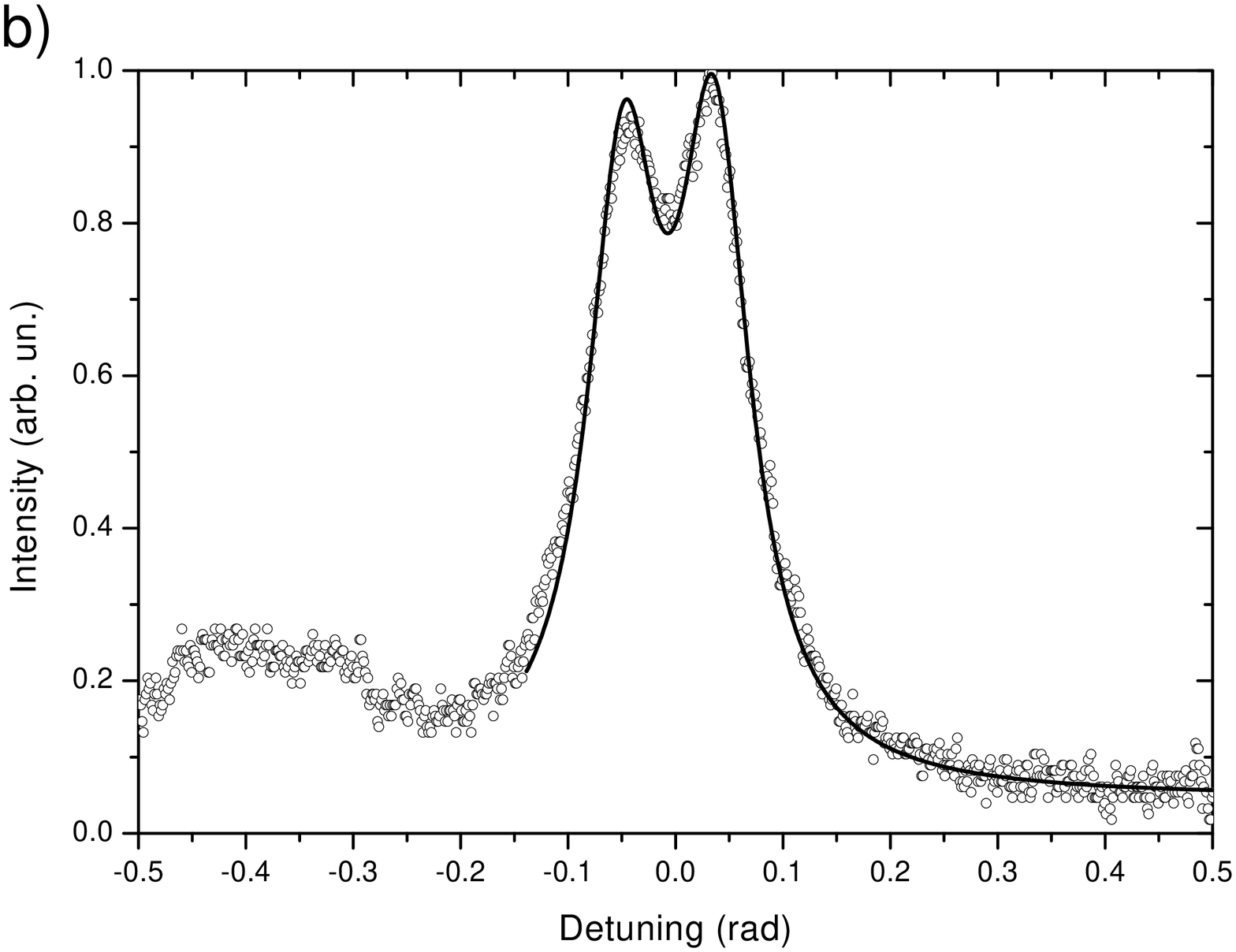}
\caption{\label{split}Cavity resonance peak for a pump LG beam with a) horizontal 
(extraordinary) polarization and b) vertical (ordinary) polarization. In the second 
case the resonance peak splits in two, clearly showing the simetry breaking 
between the two HG components of the LG beam.}
\end{figure}

In order to study this splitting, we pumped the OPO with a Laguerre-Gaussian (LG) mode 
obtained with an astigmatic mode converter \cite{cillens}. The LG 
mode is the superposition of two HG modes orthogonally oriented, that is, 
a TEM$_{01}$ and a TEM$_{10}$ mode. Once the OPO cavity is scanned, a single resonance 
peak is expected if the cavity is degenerate for the two TEM modes. Otherwise, two 
resonance peaks are expected, one corresponding to each TEM mode. 
In Fig.\ref{split}, we show the resonance peak of a high finesse cavity for a 532 nm 
LG pump. The polarization of the pump laser was rotated in order to provide 
both, the ordinary and extraordinary waves. For the vertical polarization (ordinary wave), 
a double resonance is observed as expected. This splitting shows a round trip phase difference 
of $88 mrad$, in reasonable good agreement with the predicted $93 mrad$ for 532 nm. 
On the other hand, for horizontal polarization (extraordinary wave), the LG 
resonance presents a single peak. In this case, the splitting is expected to be around 
$8 mrad$, well below the resolution of the cavity used for this measurement. 

From this analysis we conclude that the OPO can support the oscillation of an LG mode 
for the extraordinary wave, since its HG components have a degenerate (or quasi-degenerate) 
resonance frequency. On the other hand, an LG mode in the ordinary wave cannot operate 
because its HG components will not have the same resonance frequency. This explains 
the results shown in Fig.\ref{image}, that is, the orbital angular momentum (OAM) is 
transferred from the pump laser (extraordinary wave) to the idler mode (extraordinary 
wave) but not to the signal mode (ordinary wave). Notice that, under our experimental 
conditions, only one of the down converted modes oscillates in a high order transverse 
mode, while the other one oscillates in the fundamental transverse mode. 
So, the OAM exchange between pump, signal and idler modes is governed by the cavity 
dynamics under the crystal anisotropy, involving polarization and transverse profile 
aspects.

\section{Theoretical model}

Transverse multimode operation of OPOs has already been theoretically discussed 
in Ref.\cite{Schwob}. The pump beam can excite many different cavity modes for signal 
and idler, but in general it is the one with the lowest threshold that survives. 
Therefore, modes with best recovering integral should oscillate.
To extend this description to our experiment, we must take into account the walk-off 
and the astigmatism due to the crystal anisotropy. As we have seen, the astigmatism 
will introduce a phase shift between the two Hermite-Gauss components of the 
Laguerre-Gauss beam. We can choose to treat the problem either in the Laguerre-Gauss 
basis or in the Hermite-Gauss one. For the Laguerre-Gauss basis, the astigmatism 
couples the right-handed beam to the left-handed one. In the Hermite-Gauss basis, 
this coupling implies in a phase difference between the two first order modes. 
Here we chose to work in the Hermite-Gauss basis, but the change of basis is 
straightforward.

In order to study the dynamics of the relevant transverse modes, we shall 
consider the normalized mode functions 
$u_{j\,k}(x^{\prime},y^{\prime},z)$, where $j=p,s,i$ for pump, signal and idler 
respectively, and $k=0,h,v$ for the Hermite-Gauss TEM$_{00}$, TEM$_{10}$ and 
TEM$_{01}$ respectively. 
The overlap integrals, 
\begin{equation}
\Lambda_{k\,l\,m}=\int\int\int 
u_{p\,k}\,
u^*_{s\,l}\,
u^*_{i\,m}\;
dx^{\prime}\,dy^{\prime}\,dz\;,
\label{integrals}
\end{equation}
play an important role in the dynamics since they determine the transverse mode 
coupling. The mode functions $u_{j\,k}(x^{\prime},y^{\prime},z)$ are given by 
Eqs.(\ref{Zuzuy}), (\ref{Yuzuy}) and (\ref{beameq}), where astigmatism and walk-off 
effects are taken into account.
The walk-off is slightly different for pump (4.1 mrad) and idler (3.2 mrad), and 
the significant astigmatism occurs in the $z$ direction of the signal mode. 
The integrals are calculated in the whole crystal volume.

With the overlap integrals, we can obtain the dynamic equations for the transverse 
mode amplitudes. From all possible combinations of oscillating modes, the cavity 
parity will restrict the number of transverse modes for a given longitudional mode. 
If there were no anisotropic effects, with a first order Laguerre-Gaussian pump mode, 
which is odd, signal and idler must have oposite parities in order to give a nonvanishing 
overlap integral. Therefore, for isotropic propagation, if signal oscillates in a 
first order mode, idler must oscillate in the fundamental one, and vice-versa. 
In principle, this parity selection brakes down for an anisotropic medium specially due 
to walk-off. 
However, when the overlap integrals are calculated, we can see that the 
integrals for odd combinations of modes, like $(v,0,0)$ or $(v,v,v)$ for example, 
are indeed much smaller than those obtained with an even combination like $(v,v,0)$. 
This allows us to neglect many of the mode couplings and restrict the number of 
dynamic equations. Two kinds of operation regimes are observed: either the signal 
beam oscillates in the fundamental TEM$_{00}$ mode, while the idler lies in the 
TEM$_{01}$ and $_{10}$ subspace (peaks 1, 2 and 4 in Fig.\ref{peaks}), 
or the idler beam oscillates in the TEM$_{00}$ mode (peak 3 in Fig.\ref{peaks}). 
Let us describe these regimes separately.

\subsection{Signal beam operating in the TEM$_{00}$ mode}

In this case, the set of dynamic equations for pump, signal and idler 
transverse mode amplitudes is:
\begin{eqnarray}
\dot{a}_{pv}&=&-[\gamma_p+i(\Delta_p+\sigma_p)]a_{pv}-i\chi\Lambda^{*}_{v0v}a_{s0}a_{iv}
+E_{in}/\sqrt{2}
\nonumber\\
\dot{a}_{ph}&=&-[\gamma_p+i(\Delta_p-\sigma_p)]a_{ph}-i\chi\Lambda^{*}_{h0h}a_{s0}a_{ih}
-i\,E_{in}/\sqrt{2}
\nonumber\\
\dot{a}_{s0}&=&
-(\gamma+i\Delta_s)a_{s0}+i\chi\Lambda_{v0v}a_{pv}a_{iv}^*+i\chi\Lambda_{h0h}a_{ph}a_{ih}^*
\nonumber\\
\dot{a}_{iv}&=&-[\gamma+i(\Delta_i+\sigma_i)]a_{iv}+i\chi\Lambda_{v0v}a_{pv}a_{s0}^*
\nonumber\\
\dot{a}_{ih}&=&-[\gamma+i(\Delta_i-\sigma_i)]a_{ih}+i\chi\Lambda_{h0h}a_{ph}a_{s0}^*.
\label{evol1}
\end{eqnarray}
where the subindices $p$, $s$ and $i$ refer to pump, signal and idler respectively, 
and $0$, $v$ and $h$ refer to fundamental (TEM$_{00}$), vertical (TEM$_{01}$) and 
horizontal (TEM$_{10}$) transverse modes. Pump losses are described by $\gamma_p$ 
while a common decay rate $\gamma$ represents the losses for signal and idler. The 
respective cavity detunings for pump, signal and idler are $\Delta_p$, $\Delta_s$ 
and $\Delta_i$. The astigmatic symmetry breaking is accounted for through the 
frequency splitting parameters $\sigma_p$ for pump and $\sigma_i$ for idler. They 
are calculated with the help of Eq.(\ref{phase}). The pump beam amplitude transmitted 
through the input mirror is represented by the source term $E_{in}$. 
Since it is prepared in a Laguerre-Gauss mode, 
the source terms appearing in the dynamic equations for the amplitudes $a_{pv}$ and 
$a_{ph}$ are $\pi/2$ out of phase. Finally, $\chi$ is the nonlinear coupling constant.

The dynamic equations, as well as their steady state solutions, are considerably 
simplified if we express time in units of the cavity round trip time $\tau$ 
and define the following normalized variables
\begin{eqnarray}
b_{jk}&=&\chi\,\Lambda_{000}\,\tau\,\,a_{jk}\,,\;\;\;
x_{in}=\chi\,\Lambda_{000}\,\tau\,E_{in}\,,
\nonumber\\
\tilde{\gamma}_j&=&\gamma_j\,\tau\,,\;\;\;
\tilde{\Delta}_j=\Delta_j\,\tau\,,\;\;\;
\tilde{\sigma}_j=\Delta_j\,\tau\,,
\label{normaliz}\\
\eta_{k\,l\,m}&=&\frac{\Lambda_{k\,l\,m}}{\Lambda_{000}}\;.
\nonumber
\end{eqnarray}
As before, $j=p,s,i$ for pump, signal and idler respectively, and each of the 
subindices $k$, $l$ and $m$ may assume the values $0$, $h$ or $v$.
Cavity losses are around 29\% at 532nm and 1\% at 1064nm which gives 
$\tilde{\gamma}_p=145\,mrad$ and $\tilde{\gamma}=5\,mrad\,$. In the absence 
of astigmatism and walk-off the relevant normalized overlap integrals are  
$\eta_{v\,v\,0}=\eta_{h\,h\,0}=\eta_{v\,0\,v}=\eta_{h\,0\,h}=0.71$ 
approximately. When the walk-off effect is considered, the overlap integrals are 
averaged over the crystal volume. Moreover, the astigmatism is included through 
the appropriate correction of the mode functions. Taking into account the experimental 
values for the walk-off and astigmatism parameters we find 
$\eta_{v\,v\,0}=0.70\,$, 
$\eta_{h\,h\,0}=0.60\,$ and 
$\eta_{v\,0\,v} \approx \eta_{h\,0\,h} \approx 0.71\,$. 
So, a significant change is obtained only for $\eta_{h\,h\,0}$. 

It is instructive to consider the steady state solution of Eqs.(\ref{evol1}) in the 
simplified condition $\Delta_s=\Delta_i=\sigma_p=\sigma_i=0$ and 
$\eta_{v0v}=\eta_{h0h}=\eta$ which correspond to neglecting walk-off and astigmatism. 
In this case the orbital angular momentum is perfectly transferred to the idler beam 
which will also oscillate in a Laguerre-Gauss mode with the same topological charge 
of the pump beam. Therefore, the steady state solutions are
\begin{eqnarray}
I_{p-}&=&I_{i-}=0\,,\;\;\;I_{p+}=\tilde{\gamma}^2/\eta^2\,,
\nonumber\\
I_{s0}&=&I_{i+}=I_0=\frac{\tilde{\gamma}}{\eta^2}
\left[\sqrt{\frac{\eta^2\,x_{in}^2}{\tilde{\gamma}^2}-\tilde{\Delta}_p^2}\,-\,
\tilde{\gamma}_p\right]\;,
\label{sstate1}
\end{eqnarray}
where we defined the normalized intensities $I_{jk}=\left| b_{jk}\right|^2\,$. 
The Laguerre-Gauss amplitudes $b_{j\pm}$ are given in terms of the 
Hermite-Gauss amplitudes as 
\begin{equation}
b_{j\pm}=\frac{b_{jv}\pm i\,b_{jh}}{\sqrt{2}}\;.
\label{LGamp}
\end{equation}
The threshold value of $x_{in}$ for parametric oscillation is obtained by 
setting $I_0=0$ so that
\begin{equation}
x_L = \frac{\tilde{\gamma}^2}{\eta^2}
\left(\tilde{\gamma}_p^2 + \tilde{\Delta}_p^2\right)\;.
\label{lim1}
\end{equation}
As we shall see, a different threshold condition is obtained for the 
other operation regime, in which the idler beam operates in the 
TEM$_{00}$ mode.

The analytical solution for the steady state including all parameters 
is cumbersome but Eqs.({\ref{sstate1}) give us a good estimate for the 
orders of magnitude. In fact, as we discussed in section IV, the 
expected value for the pump and idler splitting parameters are ideed 
very small, $\tilde{\sigma}_p=4\,mrad$ and $\tilde{\sigma}_i=3\,mrad$ 
(the splitting parameter is half the astigmatic phase shift calculated 
from Eq.(\ref{phase})). However, this 
small splitting may be responsible for partial transfer of 
the orbital angular momentum from the pump to the idler mode. In order 
to illustrate this, we numerically integrated the dynamic equations 
(\ref{evol1}) with a fourth order Runge-Kutta method until the steady 
state was reached. In Fig.\ref{sim1} this time evolution is shown 
together with the value $I_0$ given by Eq.(\ref{sstate1}). In the 
inset, we show the expected image for signal (S) and idler (I) obtained 
with the numerical steady state results. A good qualitative agreement 
is obtained with the experimental results corresponding to peaks 
1, 2 and 4 of Fig.\ref{peaks}.
\begin{figure}[ht]
\includegraphics[clip=,width=8cm]{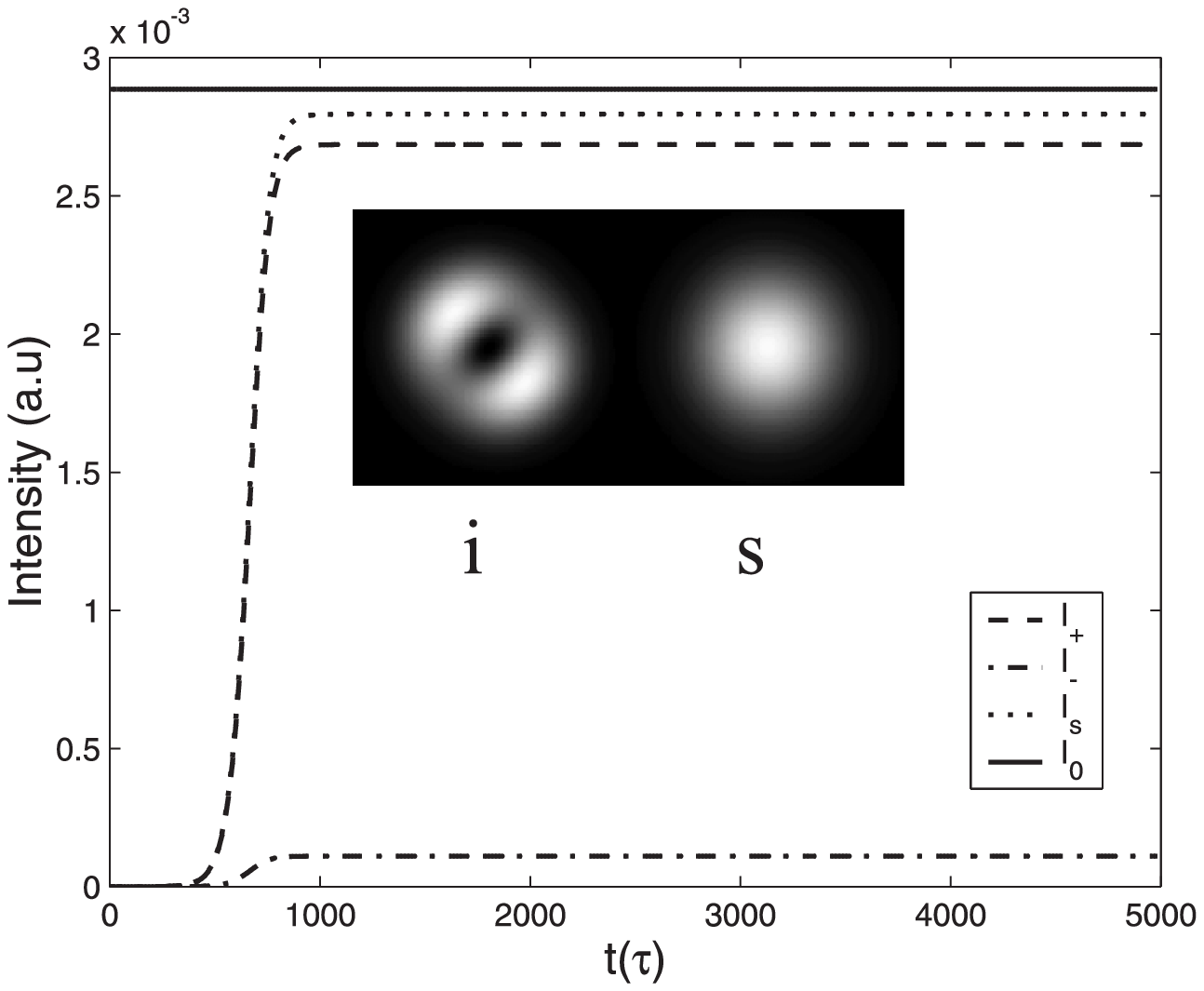}
\caption{\label{sim1}Time evolution of the down converted beam 
intensities (in units of the cavity round trip time) obtained 
from numerical integration of the dynamic equations (\ref{evol1}). 
The inset shows the corresponding expected images for signal (s) 
and idler (i). The parameter values used are $\Delta_p=0.071\,\gamma_p\,$, 
$\Delta_s=0\,$, $\Delta_i=1\,mrad\,$, $\gamma_p=145\,mrad\,$, 
$\gamma=5\,mrad\,$, $\sigma_p=4\,mrad\,$, $\sigma_i=3\,mrad\,$, 
$x_{in}=3\,x_L\,$ and $\eta_{v\,0\,v}=\eta_{h\,0\,h}=0.71\,$. 
The horizontal solid line shows the analytical value $I_0$.}
\end{figure}

\subsection{Idler beam operating in the TEM$_{00}$ mode}

In this case the dynamic equations are:
\begin{eqnarray}
\dot{a}_{pv}&=&-[\gamma_p+i(\Delta_p+\sigma_p)]a_{pv}-i\chi\Lambda^{*}_{vv0}a_{sv}a_{i0}
+E_{in}/\sqrt{2}
\nonumber\\
\dot{a}_{ph}&=&-[\gamma_p+i(\Delta_p-\sigma_p)]a_{ph}-i\chi\Lambda^{*}_{hh0}a_{sh}a_{i0}
-i\,E_{in}/\sqrt{2}
\nonumber\\
\dot{a}_{sv}&=&-[\gamma+i(\Delta_s+\sigma_s)]a_{sv}+i\chi\Lambda_{vv0}a_{pv}a_{i0}^*
\label{evol2}\\
\dot{a}_{sh}&=&-[\gamma+i(\Delta_s-\sigma_s)]a_{sh}+i\chi\Lambda_{hh0}a_{ph}a_{i0}^*
\nonumber\\
\dot{a}_{i0}&=&
-(\gamma+i\Delta_i)a_{i0}+i\chi\Lambda_{vv0}a_{pv}a_{sv}^*+i\chi\Lambda_{hh0}a_{ph}a_{sh}^*,
\nonumber
\end{eqnarray}
The transverse mode splitting now appears in the dynamic equation for the signal beam 
and is represented by the parameter $\sigma_s$. However, the splitting 
parameter is expected to be of the order of $41\,mrad$. Since cavity losses in the 
infrared are of the order of 1\%, the corresponding normalized decay rate is 
$\tilde{\gamma}=5\,mrad$, so that $\sigma_s >> \tilde{\gamma}$. Under such 
conditions it is impossible for the OPO to support the simultaneous operation of the 
$h$ and $v$ modes necessary to compose a Laguerre-Gauss mode. Therefore, the orbital 
angular momentum cannot be transferred to the down-converted beams. 
The cavity tuning will select the signal Hermite-Gauss mode whose resonance frequency 
is closer to the idler resonance. For example, for $\Delta_s=-\sigma_s$ the cavity 
frequency falls far away from the $h$ signal resonance while the $v$ mode gets on 
resonance. In this case $a_{sh}\approx 0$ and the steady state solution of 
Eqs.(\ref{evol2}) can be analytically obtained. Notice that the normalized overlap 
integral $\eta_{h\,h\,0}$ will not play any role in this case. We therefore set 
$\eta=\eta_{v\,v\,0}$ and use the same normalizations adopted in Eqs.(\ref{normaliz}) 
to find 
\begin{eqnarray}
I_{sh}&=&0\,,\;\;\;I_{ph}=\frac{x_{in}^2/2}{\tilde{\gamma}_p^2 + 
\tilde{\Delta}_p^2}\,,\;\;\;
I_{pv}=\frac{\tilde{\gamma}^2}{\eta^2}
\nonumber\\
I_{i0}&=&I_{sv}=I_0^{\prime}=\frac{\tilde{\gamma}}{\eta^2}
\left[\sqrt{\frac{\eta^2\,x_{in}^2}{2\tilde{\gamma}^2}-\tilde{\Delta}_p^2}\,-\,
\tilde{\gamma}_p\right]\;.
\label{sstate2}
\end{eqnarray}
The $h$ component of the pump beam does not couple do the down-converted modes so that 
its steady state solution corresponds just to an empty cavity. On the other hand, 
the $v$ component of the signal beam, as well as the fundamental idler mode, presents 
a steady state intensity lower than the one found in Eqs.(\ref{sstate1}) for the same 
pump level $x_{in}$. This corresponds to the situation found in peak 3 of Fig.\ref{peaks}, 
which is clearly lower than the other infrared peaks. Again, the oscillation threshold 
is readily obtained by taking $I_0^{\prime}=0$:
\begin{equation}
x_L = \frac{2\,\tilde{\gamma}^2}{\eta^2}
\left(\tilde{\gamma}_p^2 + \tilde{\Delta}_p^2\right)\;.
\label{lim2}
\end{equation} 
It is twice the threshold value for the case where the orbital angular momentum is 
transferred for the idler beam, what is also coherent with the lower height of 
peak 3 in Fig.\ref{peaks}.

The numerical evolution of Eqs.(\ref{evol2}) using a fourth order Runge-Kutta method 
was performed without the simplfying assumptions. These results are presented in 
Fig.\ref{sim2}, where the inset shows the expected images for signal and idler. The 
walk off and astigmatic effects were fully considered and a good agreement with the 
experimental result was obtained.
The theoretical model developed here will be useful for future investigation of the 
transverse mode dynamics in the quantum domain. Interesting perspectives can be 
envisaged if the OPO operation is subject to an injected signal. 
Recent studies on degenerate \cite{degen} and nondegenerate \cite{nondegen} parametric 
processes with injected signal has considered interesting issues like the preparation 
of quantum correlated states (Einstein-Podolsky-Rosen states) as well as the study of 
critical behaviours of the OPO operation \cite{critical}.
\begin{figure}[ht]
\includegraphics[clip=,width=8cm]{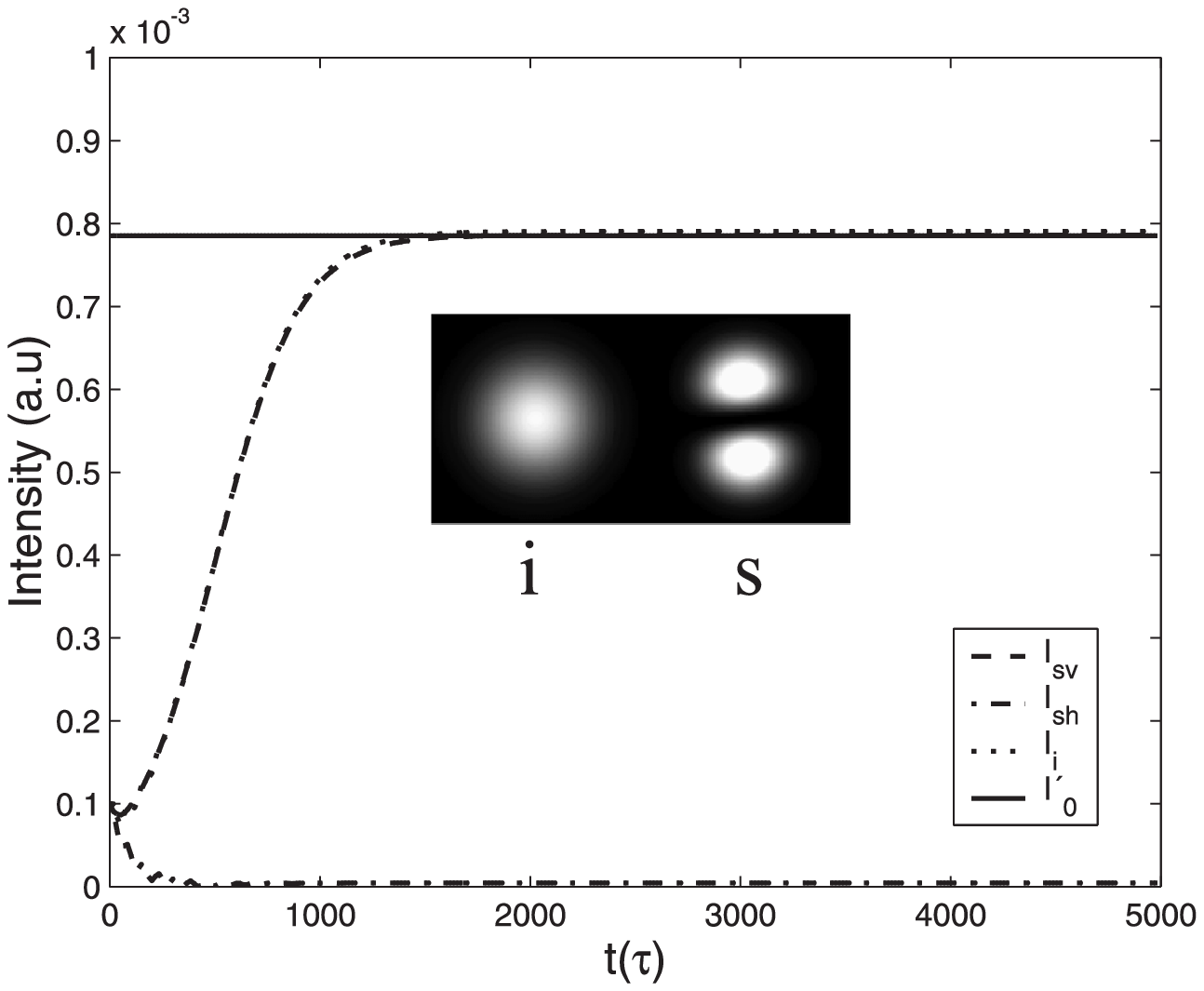}
\caption{\label{sim2}Time evolution of the down converted beam 
intensities (in units of the cavity round trip time) obtained 
from numerical integration of the dynamic equations (\ref{evol2}). 
The inset shows the corresponding expected images for signal (s) 
and idler (i). The parameter values used are $\Delta_p=0.28\,\gamma_p\,$, 
$\Delta_s=-41\,mrad\,$, $\Delta_i=0\,$, $\gamma_p=145\,mrad\,$, 
$\gamma=5\,mrad\,$, $\sigma_p=4\,mrad\,$, $\sigma_s=41\,mrad\,$, 
$x_{in}=1.5\,x_L\,$ and $\eta_{v\,v\,0}=0.70\,$. The horizontal 
solid line shows the analytical value $I_0^{\prime}$.}
\end{figure}

\section{Conclusion}

We have shown that the transfer of orbital angular momentum in intracavity parametric 
down-conversion is strongly subjected to cavity and anisotropy effects.
This can be achieved if pump, signal and idler are in a set of modes where the 
Hermite-Gauss components of the Laguerre-Gauss mode are degenerate inside the cavity. 
While that can be easily achieved for the idler beam, the signal beam  cannot fullfill 
this condition unless cavity losses are large. The signal beam can still oscillate in 
a higher order transverse mode, but in this case, the threshold power increases, and 
the orbital angular momentum is not transferred to the outcoming beams. 

Our results 
open interesting perspectives related to the quantum regime of the OPO operation. 
The experimental setup can be used to generate twin beams (intensity difference 
squeezing) and the production of the optical vortices under squeezed operation can 
lead to interesting quantum patterns in the down-converted beams. 
Moreover, the quantum dynamics of the type II operation may lead to entanglement 
between the polarization and the spatial degrees of freedom \cite{juliana}. 
We are theoretically investigating this possibility and an experiment will be 
setup in the near future.

\begin{acknowledgments}
A.Z.Khoury thanks Prof. D.Petrov for bringing the optical vortices 
to his attention during the Jorge Andr\'e Swieca Summer School on Quantum and 
Nonlinear Optics at Recife-Brazil (2000). The authors thank C.H. Monken and 
A.G.C. Moura for fruitful discussions.

The authors aknowledge partial funding from Coordena\c c\~{a}o de 
Aperfei\c coamento de Pessoal de N\' \i vel Superior 
(CAPES/PROCAD and CAPES/COFECUB projects), 
Funda\c c\~{a}o de Amparo \`{a} Pesquisa do Estado de S\~{a}o Paulo (FAPESP-BR) and 
Funda\c c\~{a}o de Amparo \`{a} Pesquisa do Estado do Rio de Janeiro (FAPERJ-BR). 
This work is mainly supported by the Conselho Nacional de 
Desenvolvimento Cient\'{\i}fico e Tecnol\'ogico (CNPq) through the 
{\textbf Instituto do Mil\^enio de Informa\c c\~ao Qu\^antica}. 
\end{acknowledgments}


\begin{thebibliography}{99}

\bibitem[*]{ca}
Corresponding author. E-mail address: khoury@if.uff.br

\bibitem{A}
R.~A.~Beth, 
Phys. Rev. \textbf{50}, 115 (1936).

\bibitem{cillens}
M. W. Beijersbergen, L. Allen, H. E. L. O. var der Veen, and J. P. Woerdman,
Opt. Comm. \textbf{96}, 123-132 (1992).

\bibitem{cillens2}
E. Abramochkin and V. Volostnikov, 
Optics Communications {\bf 83}, 123-135 (1991).

\bibitem{holog1}  
N.R. Heckenberg, R. McDuff, C.P. Smith, and A.G. White,
Optics Letters {\bf 17}, 221-223 (1992).

\bibitem{holog2}  
G.F. Brand, 
American Journal of Physics {\bf 67}, 55-60 (1999).

\bibitem{moire} J. A. O. Huguenin, B. Coutinho dos Santos, 
P. A. M. dos Santos, and A. Z. Khoury,
J. Opt. Soc. Am. \textbf{A 20}, 1883-1889 (2003).

\bibitem{amair}  
A. Mair, A. Vaziri, G. Weihs, and A. Zeilinger,
Nature (London) {\bf 412}, 313-316 (2001).

\bibitem{stimulated}
D.P Caetano, M.P. Almeida, P.H. Souto Ribeiro, 
J.A.O. Huguenin, B. Coutinho dos Santos and A.Z. Khoury, 
Physical Review A \textbf{66}, art. n$^o$ 041801 (Rapid Comm.) (2002).

\bibitem{shg1}  
K. Dholakia, N.B. Sympson, M.J. Padgett, and L. Allen, 
Physical Review A {\bf 54}, R3742-R3745 (1996).

\bibitem{shg2}  
J. Courtial, K. Dholakia, L. Allen, and M.J. Padgett, 
Physical Review A {\bf 56}, 4193-4196 (1997).

\bibitem{GattiLugiato} A. Gatti, and L. A. Lugiato, Phys. Rev. A 
\textbf{52}, 1675 (1995).

\bibitem{LugiatoGatti} L. A. Lugiato, and A. Gatti, Phys. Rev. Lett.
\textbf{70}, 3868 (1993).

\bibitem{SphericalOPO} M. Marte, H. Ritsch, K. I. Petsas, A. Gatti,
L. A. Lugiato, C. Fabre, and D. Leduc,
Optics Express \textbf{3}, 71-80 (1998).

\bibitem{LugiatoMarzoli} L. A. Lugiato, and I. Marzoli, Phys. Rev. A
\textbf{52}, 4886 (1995).

\bibitem{ConfocalSqz} L. A. Lugiato, and Ph. Grangier,
J. Opt. Soc. Am. \textbf{B 14}, 225-231 (1997).

\bibitem{Mathias} M. Vaupel, A. Ma\^\i tre, and C. Fabre,
Phys. Rev. Lett.  \textbf{83}, 5278-5281 (1999).

\bibitem{SaraDucci} S. Ducci, N. Treps, A. Ma\^\i tre, and C. Fabre,
Phys. Rev. A \textbf{64}, art. n$^o$ 023803  (2001).

\bibitem{Suret}
P. Suret, D. Derozier, M. Lefranc,J. Zemmouri, and S. Bielawski
J. Opt. Soc. Am. B \textbf{19}, 395-404 (2002).

\bibitem{vortex1} 
A. V. Mamaev, M. Saffman, and A. A. Zozulya,
Phys. Rev. Lett. \textbf{77}, 4544-4547 (1996).

\bibitem{vortex2}
Gabriel Molina-Terriza, Lluis Torner, and Dmitri V. Petrov,
Opt. Lett. \textbf{24}, 899-901 (1999).

\bibitem{graytracking}
B.~Boulanger, I.~Rousseau, J.~P. Fève, M.~Maglione, B.~Ménaert, and G.~Marnier,
IEEE Journal of Quantum Electronics \textbf{35}, 281-286 (1999).

\bibitem{Petrov}
Dmitri V. Petrov, Fernando Canal, and Lluis Torner,
Opt. Comm. \textbf{143}, 265-267 (1997).

\bibitem{Debuisschert}
T.~Debuisschert, A.~Sizmann, E.~Giacobino, and C.~Fabre, 
J. Opt. Soc. Am.~B \textbf{10},~1668-1680 (1993).

\bibitem{poincare}
M.~J.~Padgett, and J.~Courtial, 
Opt. Lett. \textbf{24},~430-432 (1999).

\bibitem{Dmitriev}
V.~G. Dmitriev, G.~G. Gurzadyan, and D.~N. Nikogosyan, {\em Handbook of
Nonlinear Optical Crystals}, vol.~64 of {\em Springer Series in Optical
Sciences}.
\newblock Berlin Heidelberg: Springer-Verlag, first ed., (1991).

\bibitem{Yariv}
A.~Yariv, {\em Quantum Electronics}.
\newblock John Wiley \& Sons, third ed. (1988).

\bibitem{Mejias}
R. Mart\'\i nez-Herrero, J. M. Movilla, and P. M. Mej\'\i as,
J. Opt. Soc. Am. A \textbf{18}, 2009-2014 (2001).

\bibitem{Ciattoni}
Alessandro Ciattoni, Gasbriella Cincotti, Damiano Provenziani, and Claudio Palma
Phys. Rev. E \textbf{66}, art. n$^o$ 036614 (2002).

\bibitem{FleckFeit}
J. A. Fleck, and M. D. Feit
J. Opt. Soc. Am. \textbf{73}, 920-926 (1983).

\bibitem{YarivYeh}
A.~Yariv and P.~Yeh, {\em Optical Waves in Crystals}.
\newblock John Wiley \& Sons, 1984.

\bibitem{Kogelnik} H. Kogelnik, T. Li; Appl. Opt. \textbf{5},
1550 (1966).

\bibitem{Monken} W. A. T. Nogueira, S. P. Walborn, S. P\'{a}dua, 
and C. H. Monken;
Phys. Rev. Lett. \textbf{86}, 4009 (2001).

\bibitem{Schwob} C. Schwob, P. F. Cohadon, C. Fabre, M. A. M.
Marte, H. Ritsch, A. Gatti, and L. Lugiato;
Appl. Phys \textbf{B 66}, 685-699 (1998).

\bibitem{xtalData} Technical data from the manufacturer (Cristal
Laser S. A.).

\bibitem{siegman} Antony E. Siegman, Lasers, University Science Books
(1986).

\bibitem{degen} M. K. Olsen, K. Dechoum, and L. I. Plimak;
Opt. Comm. \textbf{223}, 123-135 (2003).

\bibitem{nondegen} M. K. Olsen, L. I. Plimak, and A. Z. Khoury;
Opt. Comm. \textbf{215}, 101-111 (2003).

\bibitem{critical} P. Drummond, K. Dechoum, and S. Chaturvedi;
Phys. Rev. A \textbf{65}, art. n$^o$ 033806 (2002).

\bibitem{juliana}
D.P Caetano, P.H. Souto Ribeiro, J.T.C. Pardal, and A.Z. Khoury; 
Physical Review A \textbf{68}, art. n$^o$ 023805 (2003).

\end{thebibliography}
\end{document}